# Casimir-Polder forces in the presence of thermally excited surface modes

Athanasios Laliotis[1,2*], Thierry Passerat de Silans[1,2,3], Isabelle Maurin[1,2], Martial Ducloy[1,2], Daniel Bloch[1,2]

[1]*Laboratoire de Physique des Lasers, Université Paris 13, Sorbonne Paris-Cité*
[2]*CNRS, UMR 7538, 99 Avenue J-.B. Clément, F-93430 Villetaneuse, France*
[3] *Laboratório de Superfície, DF-CCEN, Cx. Postal 5086 Universidade Federal de Paraíba, 58051-900 João Pessoa, Paraíba, Brazil*

**The temperature dependence of the Casimir-Polder interaction addresses fundamental issues for understanding vacuum and thermal fluctuations. It is highly sensitive to surface waves which, in the near field, govern the thermal emission of a hot surface. Here we use optical reflection spectroscopy to monitor the atom-surface interaction between a Cs*($7D_{3/2}$) atom and a hot sapphire surface at a distance ~ 100 nm. In our experiments, that explore a large range of temperatures (500-1000K) the hot surface is at thermal equilibrium with the vacuum. The observed increase of the interaction with temperature, by up to 50 %, relies on the coupling between atomic virtual transitions in the infrared range and thermally excited surface-polariton modes. We extrapolate our findings to a broad distance range, from the isolated free atom to the short distances relevant to physical chemistry. Our work also opens the prospect of controlling atom surface interactions by engineering thermal fields.**

At the onset of the 20$^{th}$ century, the spectral behaviour of blackbody radiation, although a prototype of an incoherent source, has inspired the quantization of the electromagnetic field. It also inevitably led to the recognition of the existence of zero-point energy, the so called vacuum fluctuations. Vacuum fluctuations are ultimately responsible for the finite lifetime (spontaneous emission) and a displacement of the energy states of an isolated atom (Lamb shift). In a 'hot' vacuum (vacuum and thermal fluctuations), both lifetime and Lamb shift exhibit a temperature dependence [1,2] whose influence becomes increasingly important for ultimate optical clocks [3].

Vacuum fluctuations are also responsible for the Casimir force, an attraction between two neutral bodies such as two metallic surfaces, currently considered to be of an utmost fundamental importance while also playing a considerable role in MEMS and NEMS applications [4]. For Casimir effect studies, an exact evaluation of the temperature corrections is still required for a comparison between theory and high-precision experiments [5-8], which until now have always been performed at room temperature [9,10].

Closely related is the long-range atom-surface interaction, now generically described as Casimir-Polder (CP) effect [11,12]. It is of practical relevance for devices that trap atoms or molecules close to a surface, such as chips or nano-fibres, intended for quantum information processing experiments [13, 14]. Again, temperature corrections may play a key role, as exemplified by measurements embarked on the search of exotic Non-Newtonian gravity force [15]. Until now the only measurement demonstrating a temperature dependence of the CP attraction, was performed at distances 7-11µm, where the interaction is extremely small [16]. Critical to experimental success was an enhancement due to an out of equilibrium effect [17], where the surface is held at a higher temperature than the surrounding vacuum.

In the near field regime, thermal emission exhibits properties radically different from those of the (far-field) blackbody radiation [18], including spatial coherence [19] and a spectrum intimately related to the resonances of the material through surface modes [20]. For distances smaller than 100 nm, the CP interaction is in the van der Waals (vW) regime [11], which resembles the interaction between a fluctuating dipole and its electrostatic image in front of the surface. It is characterized by a $-C_3 \cdot z^{-3}$ potential, where z is the atom-surface distance and $C_3$ the vW coefficient [21]. In contrast to the far field regime analysed in [16,17], the near-field CP interaction strongly depends on the dielectric properties of the surface especially when surface modes coincide with atomic resonances [22]. The coupling between atoms (in the vacuum side) and surface polaritons was already exploited to demonstrate an atom-surface repulsion [23,24]. Further extension to a Förster-like real energy transfer of the atomic excitation into a surface mode was also demonstrated [25].

This resonant atom-surface enhanced process was intrinsically restricted to excited atoms at zero temperature (T=0). Conversely when surface polariton modes are thermally populated [23,26,27], the atom surface coupling can be reversed, and the vicinity with the surface can induce a virtual or real atomic excitation. With respect to the mean occupation number of photons of energy ℏω, $n(\omega, T) = \left[e^{\frac{\hbar\omega}{k_B T}} - 1\right]^{-1}$, only low-energy transitions should contribute to this type of thermal effects: at room temperature only modes of wavelength greater than 50 µm are significantly populated. This suggests that only excited high-lying state atoms or even molecules could be sensitive to temperature effects in the near field regime, while ground state atoms are typically excluded.

Here we report on measurements of the vW coefficient between a Cs*($7D_{3/2}$) atom and a sapphire surface as a function of temperature. Before discussing our experimental findings we first discuss the theoretical reasoning behind the choice of the specific atom-surface system. The vW coefficient for a given |i> level sums up the contributions of all allowed dipole couplings |i>→|j> of frequency $\omega_{ij}$, considered negative in the case of virtual emission. For our experiment a schematic diagram of the Cs energy levels and the relevant couplings is shown in Fig.1a. For an ideal conductor, one obtains simply $C_3 = \frac{1}{12}\sum_j |\langle i|D|j\rangle|^2$, with <i D | j> the dipole moment matrix elements. In the case of a real dielectric, each contribution is individually weighed by an image coefficient r($\omega_{ij}$,T), given below [27]:

$$r(\omega_{ij}, T) = \int_0^\infty S(iu) f(\omega_{ij}, T, u) du - 2Re[S(\omega_{ij})] n(\omega_{ij}, T) \qquad (1)$$

In equation (1) the first term, sometimes called 'non-resonant', resembles a distance dependent Lamb shift due to vacuum and thermal fluctuations covering the entire frequency spectrum. Here $u$ is the integration variable, S($\omega$) = [$\varepsilon(\omega)$ -1] /[$\varepsilon(\omega)$ + 1] is the surface response (see Fig1b for sapphire), with $\varepsilon(\omega)$ the complex bulk permittivity of the window and $f(\omega_{ij}, T, u)$ is a generic function that groups all other contributions and turns the integral into a discreet sum over the Matsubara frequencies, when T≠0. The second term, often called "resonant", is reminiscent of the interaction between a classical oscillating dipole and its own reflected field [28,29]. For virtual absorption ($\omega_{ij}$ >0), it is simply proportional to the surface response at the atomic frequency and the mean photon occupation number n($\omega_{ij}$>0,T), hence null at T=0 [22]. For a virtual emission ($\omega_{ij}$ <0) the sign is inverted and one has to account for one additional photon due to spontaneous emission, so that $n(\omega_{ij} < 0, T) = -[1 + n(|\omega_{ij}|, T)]$.
In Fig1c, we show the image coefficient of sapphire for a virtual absorption at different temperatures. A large temperature dependence is observed for virtual transitions whose frequency is in the vicinity of the surface polariton at 12.1 μm (830 cm$^{-1}$). For smaller frequencies the two terms of eq. (1) tend to compensate (inset of Fig.1c) and temperature effects are smaller. This compensation is perfect only in the case of a dispersion-less ideal conductor [30]. Interestingly, in the high frequency side of the spectrum a small, but non-negligible temperature dependence is observed despite the small number of average photons present. This can be traced back to sapphire dispersion as well [31]. In the case of virtual emission the temperature dependence of the image coefficient is analogous and adds to the resonant behaviour at T=0 mentioned above. The table of Fig.1d shows the contributions of each dipole coupling relevant for Cs ($7D_{3/2}$). Remarkably, the $7D_{3/2} \rightarrow 5F_{5/2}$ virtual absorption at 10.8 μm, responsible for only half of the $C_3$ value at low temperature, carries nearly all of the $C_3$ temperature dependence. Conversely, the two important virtual atomic emissions $7D_{3/2} \rightarrow 8P_{1/2}$ at 29 μm and $7D_{3/2} \rightarrow 8P_{3/2}$ at 39 μm, exhibit only small temperature dependence despite their small energy.

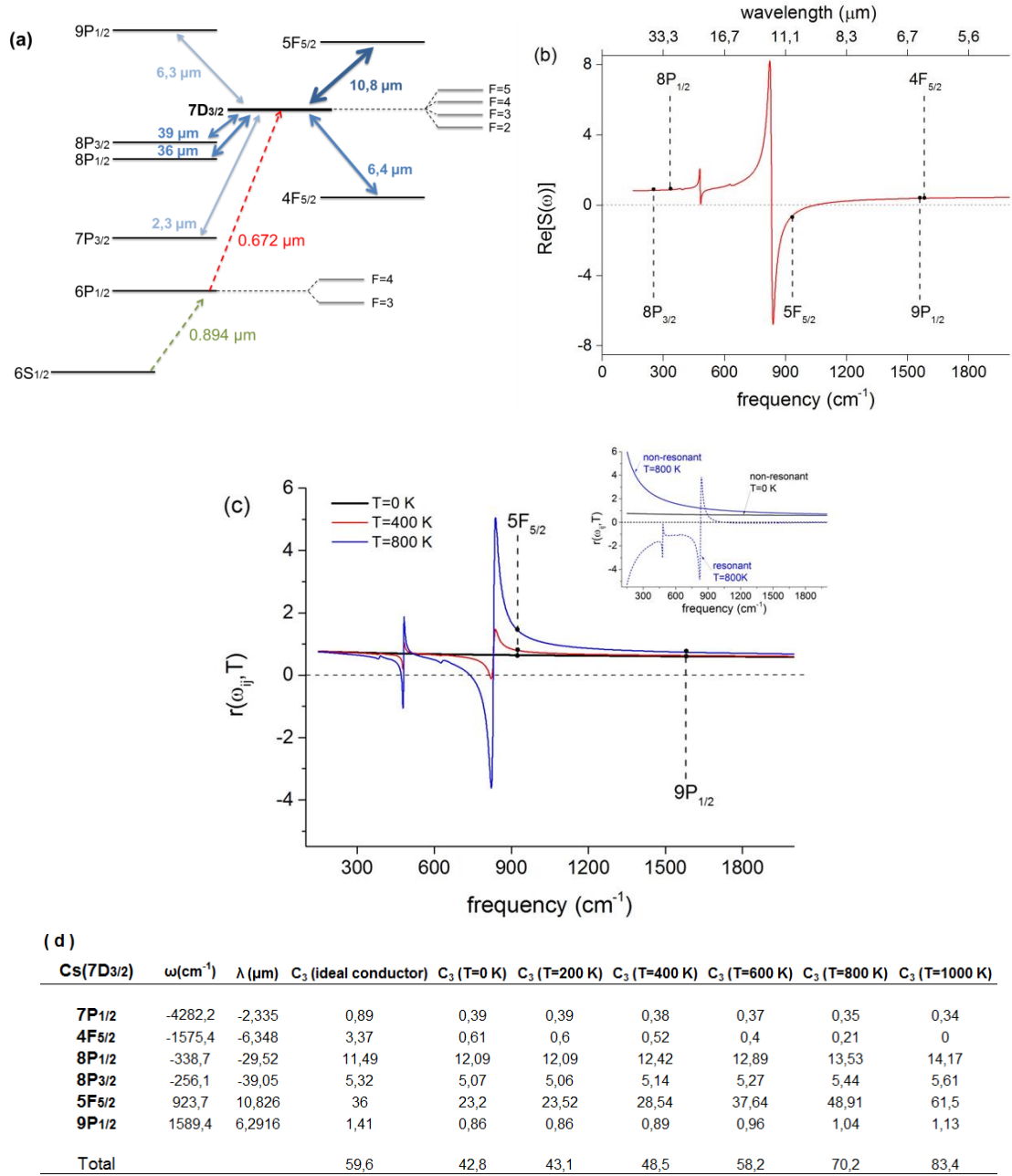

**Figure 1:** (a) The Cs energy levels and the dipole transitions relevant to our experiment. The dashed arrows represent the optical excitation scheme, with pumping from the ground $6S_{1/2}$ and probing from the $6P_{1/2}$ level. (b) The real part of the sapphire surface response, $Re[S(\omega)]$. We use experimental measurements performed at room temperature given in [41]. The solid points show the positions of the relevant dipole coupling from $7D_{3/2}$. (c) The image coefficient of a sapphire surface for a virtual atomic absorption ($\omega_{ij}>0$) as a function of the transition frequency for T=0 K, T=400 K, T=800 K. The inset shows separate plots of resonant (dashed) and non-resonant (solid) contributions for T=800 K and for T=0 (resonant term is zero). One notes that the monotone behaviour predicted at T=0 disappears at finite temperatures. (d) Table shows the values of the global $C_3$ coefficient, in kHz $\mu m^3$, as well as the individual contributions of the most important dipole couplings for different temperatures. The $C_3$ values for an ideal conductor are also given. Here we use the values of the dielectric constant as measured at T=300 K, ignoring the effects of temperature on the dielectric properties of the material. These are analysed in detail in [39] but in our case they are negligible.

For our experimental demonstration, we use selective reflection (SR) spectroscopy on the $6P_{1/2} \rightarrow 7D_{3/2}$ line (λ = 672 nm), that requires a prior pumping step $6S_{1/2} \rightarrow 6P_{1/2}$ ($\lambda_p$=894 nm). Linear SR spectroscopy on the interface between a dielectric and an atomic vapour is a long established spectroscopic method, in which the changes of the reflectivity coefficient near the atomic resonance yield information on the vapour spectrum close to the surface, typically probing a λ/2π depth (~100 nm) largely compatible with the vW regime. SR spectroscopy, in its frequency modulated (FM) version, has been rendered suitable for an accurate evaluation of the vW–induced transition shift through numerous successive developments [21,23,24,32-35]. Here, with the prior pumping to a resonant level, our experimental set-up (Fig.2a) is similar to the one previously used to detect surface repulsion [23,24]. Our experiment essentially measures the temperature dependence of the interaction between sapphire and Cs$^*$($7D_{3/2}$), as the attraction exerted on Cs$^*$($6P_{1/2}$) is negligible [35].

Our measurements are conducted in an all-sapphire Cs vapour cell, shown in Fig.2b. Sapphire has the benefit of being a robust material, reliable enough to allow the construction of a high-temperature cell [36] that operates at temperatures exceeding 1000 K. The window, on which the SR experiment is performed, is "super-polished" and post-annealed with an average roughness of ~ 0.3 nm, measured by atomic force microscopy. A differential heating system allows an independent control of the Cs vapour density which is governed by a low temperature reservoir usually operated in the range of ~400-450 K. A separate oven controls the temperature of the window which is in equilibrium with its surrounding environment. This temperature T, varied between 500-1000 K, is the one relevant for the atom-surface interaction. Experimentally, we scan the frequency of the 672 nm laser, sent under a near normal incidence on the hot sapphire window, around one of the ensemble of resolved hyperfine components, *i.e* either the $6P_{1/2}$ (F=4) → $7D_{3/2}$ {F'=3,4,5}, or the $6P_{1/2}$ (F=3) → $7D_{3/2}$ {F'=2,3,4} transitions (see Fig.1a). An auxiliary saturated absorption (SA) experiment is implemented (Fig. 2a) to provide frequency markers of the free-atom spectrum. As in [23, 24, 33-35], and thanks to the linearity of SR spectroscopy, a universal theoretical model [32], taking into account the spatially inhomogeneous vW shift of the transition, the transient regime of optical interaction for moving atoms, and the atomic velocity distribution, is used to fit the recorded spectra. The adjustable parameters are the vW coefficient ($C_3$) and the linewidth of the atomic transition (Γ).

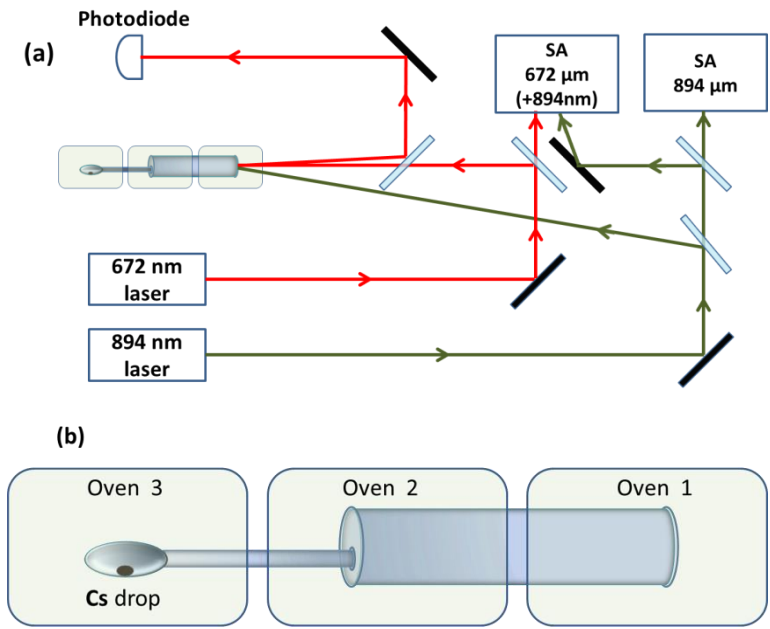

**Figure 2**: (a) Schematic of the experimental set-up. The essential experiment is a FM selective reflection at normal incidence at 672 nm, the oblique irradiation at 894 nm beam pumps the atoms to the intermediate $6P_{1/2}$ level; the two saturated absorption experiments, at 894 nm ($6S_{1/2}$→$6P_{1/2}$), and at 672 nm ($6P_{1/2}$→$7D_{3/2}$), are respectively used for the locking and monitoring of the laser frequencies. (b) Schematic of the heated Cs cell with sapphire windows showing the differential heating, allowing separate control of the atomic vapour density, and of the window temperature.

In Fig.3a we present experimental spectra measured at the lowest (T=490 K) and the highest (T=1000 K) window temperature. One notes the tremendous influence of the atom-surface interaction (see inset for $C_3 = 0$), and important differences in the spectra when changing temperature. The theoretical fits are in such good agreement that they are hardly visible.

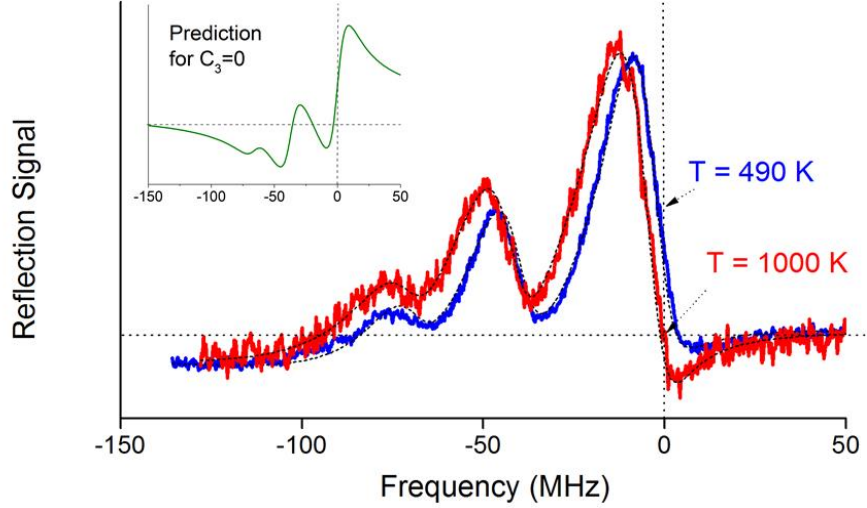

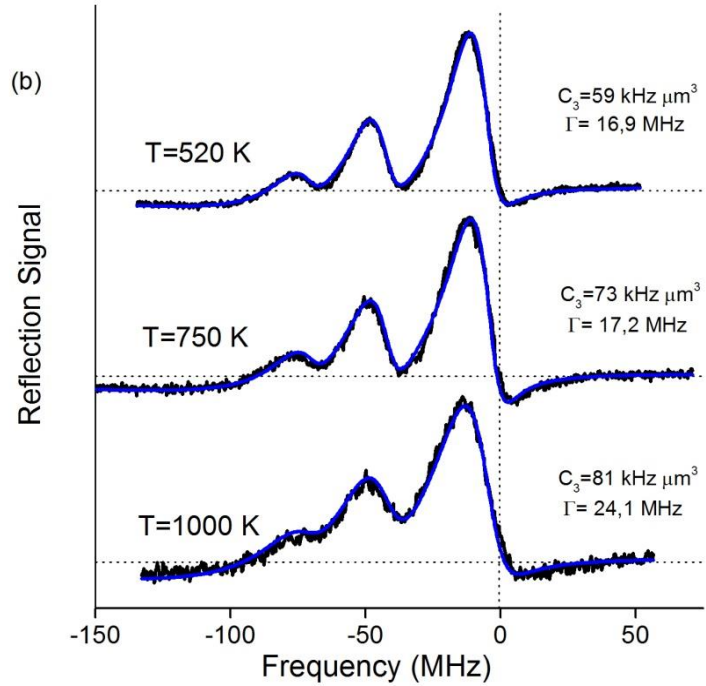

**Figure 3:** (a) Normalised FM SR spectra on the $6P_{1/2}$ (F = 4) → $7D_{3/2}$ (F' = 3, 4, 5) transition (672 nm) for a window temperature at T = 490 K (blue) and at T = 1000 K (red). The frequency axis is referenced to the hyperfine component of the $6P_{1/2}$ (F=4) →$7D_{3/2}$ (F'=5) transition as determined by the auxiliary saturated absorption. In the inset we show the predicted FM SR spectrum in the absence of atom-surface interaction, *i.e* $C_3$=0. The observed shift and lineshape distortion of the experimental curves are a clear evidence of the atom-surface interaction. The lower signal/noise ratio for 1000 K window is further discussed in the Methods section. The dashed black lines are the respective theoretical fits demonstrating the same linewidth for both temperatures Γ=19 MHz and an increase of $C_3$ with temperature from 55 kHzµm³ to 86 kHzµm³ (b) Experimental spectra (black) for FM selective reflection at the F=4 → F'={3,4,5} hyperfine manifold for various window temperatures along with the corresponding theoretical fits (blue) The parameters used to fit the spectra are also indicated.

The linewidth of the atomic transition is kept constant in both experiments so the changes in the shape of the spectra can be traced directly back to the change of the $C_3$ coefficient. Fig. 3b presents several other spectra, for three indicative values of the window temperature (T=520 K, T=750 K, T=1000 K). Excellent fittings are observed in all cases, and the evolution of the fitting parameters demonstrates a clear increase of the strength of the atom-surface interaction with temperature. We have performed various independent tests to check the consistency of our measurements, in the same way as in [23]. For a given window temperature, we notably vary the Cs vapour pressure by changing the reservoir temperature, therefore changing the value of $\Gamma$ due to collision broadening. These changes, although affecting the amplitude and shape of our spectra, do not modify the fitted $C_3$ values. Additionally, when comparing the two different hyperfine manifolds, one observes strong differences in the overall shape of the curves owing to the different relative weight of each hyperfine transition, but identical values for $C_3$ and $\Gamma$ are extracted for both spectra. Our measurements were also performed for many different spots on the sapphire window, with no significant changes. This is a crucial point because local structure variations (*e.g.* roughness) or impurities (defects or ions) may severely affect atom-surface interaction and bias our measurements [37,38]. Consistent measurements were even found for a set of preliminary measurements performed in a sapphire cell with a lower surface quality. These precautions also assure us that no chemical degradation of the surface occurs with continuous exposure on chemically aggressive alkali vapour at elevated temperatures. Previous efforts to look for a thermally induced repulsion predicted to occur at moderate temperatures for Cs*(8P) [34,35,39] coupled to fluoride windows in the 30-40 μm range had turned out to be unsuccessful [40] due to the extreme mechanical and thermal fragility of fluoride windows. On the contrary, sapphire has been remarkably stable under these extreme conditions.

Our experimental results are summarised in Fig. 4. The experimental error bar for $C_3$ is on the order of 15% when fitting a single spectrum [40], but reduces down to 5-10% after using all our independent measurements. One observes a significant increase of the vW coefficient with temperature (~50 % from 500K to 1000 K), which is in excellent agreement with the theoretical prediction, unambiguously demonstrating the fundamental thermal effects on the CP interaction. Our prediction requires the knowledge of the transition probabilities [33] and of the dielectric constant of sapphire on the entire spectrum. This introduces some small uncertainty in the theoretical values. The major contributing factor comes from the transition probabilities which are not accurately known for all dipole couplings. We use theoretical estimates given in [40] to which we moderately assign a 5% uncertainty. In Fig. 4 we also show theoretical predictions using two different sets of experimental measurements [39,42] of the sapphire dielectric constant. The former [39] analyses in detail the effects of temperature (up to 800 K) on the dielectric constant of sapphire. The latter [42] also provides data for the sapphire dielectric constant on the extra-ordinary axis, which in our experiment is perpendicular to the window. Taking sapphire birefringence into account [43] has a small effect on our theoretical predictions. Further calculations show that the atom surface interaction is independent of the actual axis orientation of sapphire to within 1%. Note that the uncertainties of the dielectric constant of sapphire may have a significant impact if a virtual transition coincides with surface polariton frequency.

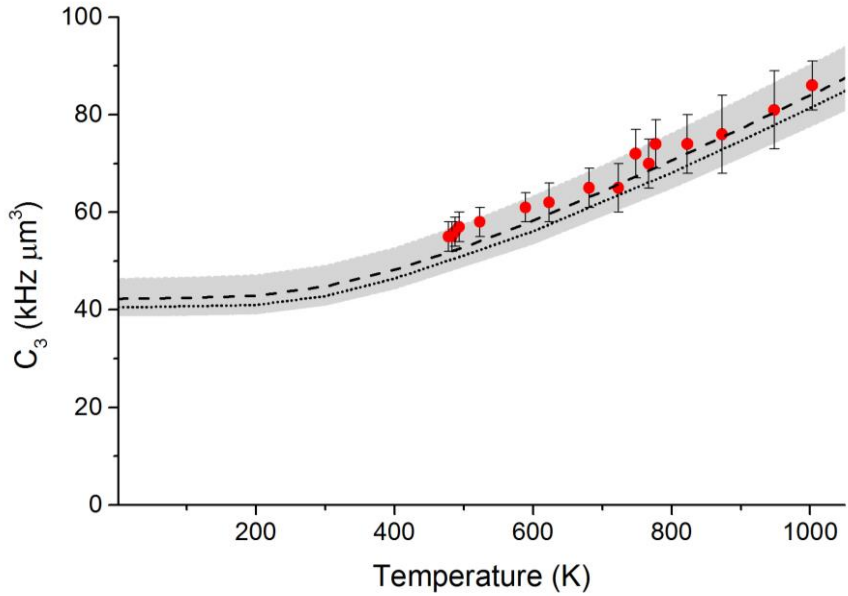

**Figure 4:** The $C_3$ (van der Waals) coefficient as a function of temperature: red points represent experimental measurements. The dashed and dotted lines respectively represent theoretical predictions using the dielectric constant of sapphire reported in [39] and in [42]. The grey shaded area of 5% plotted around the theoretical curves represents the uncertainty assigned to the transition probabilities.

To extrapolate our findings to a larger scope and a variety of regimes, we also considered the effects of propagation for our $Cs^*(7D_{3/2})$/sapphire system. Fig. 5 shows for various temperatures the spatial evolution of the atom-surface interaction potential, as determined through a full QED calculation. It confirms the validity of the $-C_3z^{-3}$ vW limit at any temperature for distances up to 100-200 nm, typically probed by SR spectroscopy. The typical evolution of the shift of a ground state atom, from $z^{-3}$ (vW) to $z^{-4}$ (CP asymptotic limit), and then after distances greater than the 'thermal length' back to $z^{-3}$ (Lifsitz) [16,17], is unnoticeable. Rather, the major correction to the vW limit is an oscillating behaviour from attraction to repulsion with an envelope decreasing slower than $z^{-3}$, typical of an excited atom due to the reflection of spontaneous emission [28]. In particular, the major contribution to oscillations comes from the $7D_{3/2} \rightarrow 7P_{3/2}$ transition at 2,33 μm for which propagation effects 'switch on' much closer to the surface. Temperature dependence is small in this distance regime due to the absence of thermal photons at this wavelength. Thermal effects are the strongest in the near field as a consequence of the surface polariton. We point out here that our calculations give the free energy shift (see [27] and refs therein) due to the presence of the hot surface, ignoring the effects of Black Body Radiation (BBR) on an isolated atom. Taking into account this effet, discussed in detail in [1], the thermal contribution to the CP shift is dominant only when the atom is less than 1 μm away from the surface.

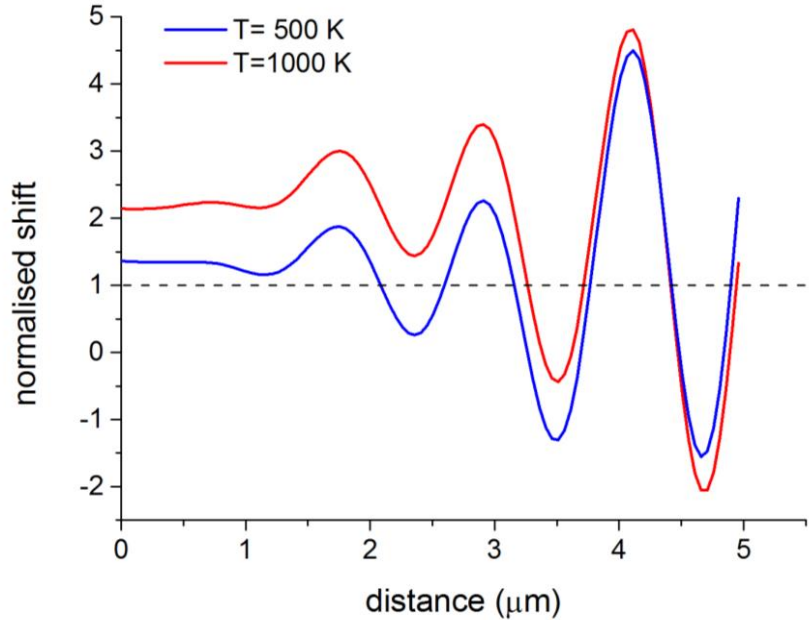

**Figure 5:** Spatial dependence of the free energy shift of the $Cs(7D_{3/2})$ atom in front of a sapphire surface, for T= 500K (blue) and 1000K (red), normalized to the spatially varying van der Waals electrostatic limit, $-C_3(T=0)/z^3$ (dashed black line).

The distance limit that our experiments explore extends down to the nanometric range (~1 nm). This distance is already in the asymptotic "long-distance" limit for various physical chemistry processes, of interest for *a priori* calculations [44,45]. Our result demonstrating that the CP attraction increases with temperature contradicts the generally accepted behaviour of thermal desorption. This should, nonetheless, be taken with a grain of salt since the vW potential measured here is for an atomic excited state. Our fundamental studies can, however, provide new insights for the microscopic understanding of physico-chemical processes. As an example, in a temperature dependent desorption process the thermally excited surface modes may selectively enhance a channel of energy transfer thus favouring a particular energy state for the desorbed atom or molecule.

In conclusion we have established that the interaction between a $Cs^*(7D_{3/2})$ atom close to a hot sapphire surface, in thermal equilibrium with the environment, increases with temperature. This effect relies on the coupling of atomic excitations to surface polariton modes, which is critical in the vW regime of distances. A very good quantitative agreement with theory is observed. The influence of temperature reveals the QED nature of the van der Waals interaction and may provide a useful knob to control forces between neutral atoms and surfaces up to their complete cancellation or change of sign. It is in particular possible to extend, with temperature as an extra-parameter, the current studies on the Casimir interaction with specially designed nano- or micro-structured surfaces, where the shaping allows tailoring the thermal field and possibly the interaction (repulsion, torque). Atoms may also be used as sensitive quantum probes of coherent thermal fields [19] or even of the correlated properties of thermal emission [46,47]. An alternative geometry would be a thin vapour cell [38, 48], which is a simple realization of a dielectric cavity, and an example of a situation where the temperature-dependence of surface interaction should depend on the surface geometry [49].

**Acknowledgements**

It is a pleasure to thank J.R. Rios Leite for fruitful and fascinating discussions as well as for his long standing interest in our work, M-P Gorza, P. Chaves de Souza Segundo and H. Failache for early contributions to this work and numerous discussions, D. Sarkisyan for the fabrication of high temperature cells, and G. Pichler for kindly lending us an sapphire vapour cell. The France-Brazil co-operation has been supported by CAPES-COFECUB Ph 740/12 and 456-04.

**METHODS**

**- Cell , ovens, and equilibrium temperature :**

A schematic of the specially designed Cs cell, prepared by the group of D. Sarkisyan (Ashtarak, Armenia) is shown in Fig2(b). It consists of a 10 cm long cylindrical sapphire tube onto which two sapphire windows are glued. The primary window is "super-polished" with the *c*-axis perpendicular to the surface. It has a small wedge allowing us to select the reflection from the vapour interface. The mineral gluing is capable to resist temperatures up to 1200 K. The secondary window is of a lower surface quality. The 7cm long sidearm, parallel to the cell main body, is glued on a small hole drilled on the secondary window.

For this sidearm Cs cell, we have implemented a multiple section oven around the whole cell. Oven 1, about 5 cm long, controls the temperature of primary window region. It keeps the sapphire window in thermal equilibrium with the surrounding blackbody radiation. The temperature inside oven 1 can be chosen in the 500-1000 K range. For such temperatures, the Cs vapour pressure would be considerable, and pressure broadening too high for any meaningful SR measurements. Ovens 2 and 3, also about 5 cm long each, impose a large temperature gradient on the entire cell and keep the sidearm as a "cold spot" of the cell, where Cs condenses in liquid form. Oven 3 controls the temperature of the sidearm, and therefore the Cs vapour pressure inside the cell. It is typically kept at temperatures around 400-450 K corresponding to a Cs equilibrium density ~ $10^{14}$ - $10^{15}$ at/$cm^3$.

The temperature of each oven is measured with thermocouples, whose location can be moved, in order to evaluate the temperature homogeneity of the oven. Our readings inside oven 1 confirm that temperature is homogeneous to within 10-20 K. It is important to notice that the implementation of the differential heating in the Cs cell generates a temperature gradient on a length scale (~ 10 cm) which does not impact the temperature relevant for the atom surface interaction. In addition, our measurements probe an atom-surface interaction at the level of a single atom, making the "temperature" of the atomic vapour, as an ensemble of moving atoms, irrelevant for our investigation.

**Atomic vapour and optical pumping**:
For our SR measurements, Cs atoms are first pumped to the $6P_{1/2}$ level, by an 894 nm pump laser. The role of this pump laser is to provide a quasi-thermal velocity distribution of $Cs^*(6P_{1/2})$ atoms at the Cs vapour interface. This was already achieved in previous experiments [23, 24]. We benefit from collisions, which redistribute the excitation to all velocities of both hyperfine components. To avoid any velocity-selective pumping the pump laser frequency is resonant to the $6P_{1/2}$ hyperfine level which is not directly probed by the 672 m SR laser (*e.g.* pump reaching the $6P_{1/2}$ (F=4) level, and probe starting from the $6P_{1/2}$ (F=3) level, or vice-versa).

The cell structure and multiple oven design means that the atomic gas is not in equilibrium. Because of collisions, one expects a uniform pressure throughout the cell and a Cs velocity distribution that depends primarily on the local surface temperature. This implies that the Cs density is locally smaller in the hottest areas of the cell, while simultaneously the atomic velocity distribution gets broader. As a result the SR signal reduces when increasing the temperature of the primary window. This is clearly observed in our measurements (Fig.3a).

**Optical set-up**

A DBR laser at 894 nm is used to pump the atoms to the $6P_{1/2}$ intermediate level. The pumping beam is relatively high power ~10 mW with a waist of ~800 μm, and can be sent to the cell on an oblique incidence, practically limited by the geometry of oven 1. An extra on/off amplitude modulation (AM) is applied to the pump beam with an acousto-optic modulator, at a frequency of 10 kHz, in order to improve detection sensitivity. The pump frequency is locked through an auxiliary set-up of saturated absorption at 894 nm.

The SR probe is an extended cavity laser at 672 nm. The probe beam is slightly smaller ~ 500 μm waist, to ensure a complete overlap with the pump, and less powerful, ~50 μW, to avoid saturating the atoms. It is sent to the sapphire window under nearly-normal incidence (on the order of 10 mrad). The laser frequency is scanned by applying a voltage on the piezoelectric element attached to the external grating. The laser is frequency modulated (FM) at 1 kHz, with ~5 MHz amplitude. An auxiliary saturated absorption experiment and a low finesse Fabry Perot cavity provide us with frequency markers. An improved method of controlling the frequency scan by beating the probe with a frequency locked reference laser was also used without any observable improvement in the quality of our SR spectra fits. The reflection of the probe beam is measured with a low-noise photodiode and then demodulated with two cascaded lock-in amplifiers. The atomic signal is typically between four and five orders of magnitude smaller than the reflection from the sapphire/vapour interface.

**Fitting method**

$C_3$ measurements are extracted by fitting experimental spectra to a library of theoretical curves depending upon a dimensionless parameter [$A = 2C_3\Gamma^{-1} (2\pi/\lambda)^3$, with $\Gamma$ the optical linewidth]. Linearity of SR spectroscopy imposes the relative amplitudes and the frequency spacing of the hyperfine components of a given manifold to their theoretical values. The experimental spectrum is compared to a theoretical curve, of a given A, by dilating the frequency axis, in order to obtain $\Gamma$ [23, 24, 32-35]. We also allow for an overall offset of the spectrum and adjustment of the global amplitude of the theoretical curve in order to reflect experimental instabilities. We then change the dimensionless A value, and repeat the process until the best fit is identified. This provides a measurement of the $C_3$ coefficient and the transition linewidth $\Gamma$ [23, 24, 32-35]. The error bars are chosen by identifying the values of $C_3$ and $\Gamma$, for which the fit curve falls clearly outside the bounds defined by the noise of the experimental curve. Our fitting method may allow for a global pressure shift, which appears negligible when fitting our data. We can also account for a finite Doppler width correction [24], as the theoretical modelling is essentially developed in the infinite Doppler width approximation. This correction mostly modifies the wings of the SR spectrum. Here, we can safely ignore this correction (attempted for some fittings with no significant changes). This is probably because the width of the hyperfine manifold, which determines the overall lineshape, is much smaller than the Doppler width.